\documentclass[twocolumn,showpacs,preprintnumbers,amsmath,amssymb,final]{revtex4}

\usepackage{graphicx}
\usepackage{dcolumn}
\usepackage{bm}

\def\BEq{\begin{equation}}
\def\EEq{\end{equation}}
\def\BEqA{\begin{eqnarray}}
\def\EEqA{\end{eqnarray}}
\def\BEn{\begin{enumerate}}
\def\EEn{\end{enumerate}}
\def\BWT{\begin{widetext}}
\def\EWT{\end{widetext}}
\def\a{\alpha}

\def\b{\beta}

\def\bra{\langle}

\def\ket{\rangle}

\begin{document}


\title{
Interaction-assisted SU(2) adiabatic holonomies in  
Josephson phase qubits}

\author{Andrei Galiautdinov}
 \email{ag@physast.uga.edu}
\affiliation{
Department of Physics and Astronomy,
University of Georgia, Athens, Georgia
30602, USA
}

\date{\today}

\begin{abstract}

We propose a scheme for generating SU(2) adiabatic geometric phases in a 
circuit consisting of 
three capacitively coupled flux-biased Josephson phase qubits.

\end{abstract}


\maketitle


\section{Introduction}

\label{sec:INTRO} 

When slowly changing, external controls that govern adiabatic evolution of a 
quantum system 
return to their original values,
the phase of system's wave function acquires a purely geometric contribution
independent of both system's energy and duration of the adiabatic process.
This $U(1)$ contribution, initially discovered by Berry \cite{BERRY1984}, 
mathematically interpreted by 
Simon \cite{SIMON1983}, and later generalized to non-abelian $U(N)$ operations  
by Wilczek and Zee \cite{WILCZEK&ZEE1984}, has been the source of many 
interesting developments in various fields of chemistry and physics 
\cite{geomPhaseLiterature}, such as molecular and spin dynamics, optics, 
fractional quantum Hall effect, various branches of condensed matter 
physics \cite{geomPhaseBooks}, and others, such as, for example, 
the field of quantum information processing \cite{HQC}.

Of particular interest to us is the possibility of 
observing {\it non-abelian} geometric phases (SU(2), in our case) in 
superconducting circuits with Josephson junctions \cite{nonAbelianScenarios}. 
Such circuits are currently considered as promising candidates for scalable 
quantum computing architectures \cite{YOU_review05}. Even though in this 
paper we consider a definite physical system (the so-called capacitively 
coupled flux biased Josephson phase qubits \cite{MARTINIS2007}), our analysis 
may apply equally well to any quantum computing architecture whose dynamics 
is described by the Hamiltonian given in Eq. (\ref{eq:Hint_2couplers}). This very 
simple Hamiltonian has properties that make 
the corresponding system ideally suitable for actual realization of non-abelian 
geometric phases. The most important of these is the presence of two doubly 
degenerate subspaces, one of which, $V_1$, {\it can} support non-abelian phase, 
and the other, $V_2$, {\it cannot}. Experimentally then, $V_2$ may be used as 
fiducial, reference subspace relative to which various interferometric experiments 
involving $V_1$ could be performed. 

To recall how geometric phases appear in quantum mechanics and to fix our 
notation, 
let us consider a quantum system whose Hamiltonian depends on the time $t$ 
via 
several controllable parameters $\lambda(t)\equiv \{\lambda^1(t), \lambda^2(t), 
\lambda^3(t), \dots\}$, ${\cal H}={\cal H}(\lambda(t))$. It is assumed that 
${\cal H}(\lambda(t))$ forms an iso-degenerate family of Hamiltonians
(no level crossings as $\lambda$ varies) \cite{HQC}. Let the instantaneous 
energy eigenbasis $\chi_{a} (\lambda(t))$ within a give degenerate subspace 
be chosen,
\BEq
\label{eq:Hat}
{\cal H}(\lambda(t))\chi_{a}(\lambda(t))=E(\lambda(t))\chi_{a}(\lambda(t)), \;
a = 1,2,\dots, N.
\EEq
At $t=0$, prepare the system in one of the energy eigenstates,
$\psi_{a}(0)\equiv\chi_{a}(\a(0))$, and let it evolve according to the 
Schr\"{o}dinger equation,
\BEq
\label{eq:SCHRODpsi}
id\psi_{a}(t)/dt = {\cal H}(\lambda(t))\psi_{a}(t).
\EEq
Then, in accordance with the adiabatic theorem, at some later time $t$,
the state $\psi_{a}(t)$ will be a linear superposition of various 
$\chi_{b}(\lambda(t))$ belonging to the same degenerate subspace. Therefore, 
in general,
\BEq
\label{eq:psi}
\psi_{a}(t) \approx e^{-i\int_0^t E(\lambda(t'))dt'} \sum_b \chi_{b}(\lambda(t))
U_{ba}(\lambda(t)),
\EEq
where the first factor on the right is the usual dynamical phase, and 
$U(\lambda(t))$ is the unitary matrix representing additional, purely 
geometric ``rotation.'' The $U(\lambda(t))$ is the sought for non-abelian 
geometric phase. It is given by the path-ordered exponential,
\BEq
U(\lambda(t)) = {\cal P} \exp\left\{-\int_0^t \sum_i {\cal A}_{i}(\lambda(t')) 
\; \dot{\lambda}^i(t') \;dt' \right\},
\EEq
where
\BEq
\label{eq:ADIABATICCONNECTION}
\left[{\cal A}_{i}(\lambda)\right]_{cb} \equiv 
\bra \chi_{c}(\lambda)|\partial/\partial\lambda^i|\chi_{b}(\lambda)\ket
\EEq
is the so-called {\it adiabatic connection}, also known as the $u(N)$-valued 
gauge potential. 
In general, when $\lambda(t)$ traces a closed loop ${\cal C}$ in the 
parameter space, the 
system picks up a nontrivial geometric phase (the {\it holonomy}),
\BEq
U_{{\cal C}} = {\cal P} \exp\left\{-\oint \sum_i {\cal A}_{i}(\lambda)
d\lambda^i \right\}\neq 0,
\EEq
independent of the speed with which the loop had actually been traversed 
(as long as it was done adiabatically). In the next section we describe a 
multi-qubit solid state quantum computing architecture capable of supporting 
such a phase. 

\section{Capacitively coupled tripartite system and its holonomies}

One system that leads to nontrivial, SU(2) holonomies consists 
of three capacitively coupled Josephson phase qubits \cite {MARTINIS2007}
whose effective 
Hamiltonian
in the rotating wave approximation is given by (cf. \cite{maxEnt2008})
\BEqA
\label{eq:Hint_2couplers}
{\cal H}(B_x,B_y) &=& (1/2)
[ B_x(\sigma^x_1+\sigma^x_2)+B_y(\sigma^y_1+\sigma^y_2)
\nonumber \\
&& +
g\left( 
\sigma^x_3\sigma^x_2+\sigma^y_3 \sigma^y_2
+
\sigma^x_3\sigma^x_1+\sigma^y_3 \sigma^y_1
\right)
\nonumber \\
&& + 
J\left( \sigma^x_2\sigma^x_{1}+\sigma^y_2 \sigma^y_{1}\right)
] ,
 \EEqA
where, in polar coordinates,
$B_{x} = B \cos \phi$,
$B_{y} = B \sin \phi $.
Here, qubits 1 and 2 experience the same magnetic field produced by 
identical microwave driving on both
qubits. This microwave drive is the parameter $\lambda \equiv \{B, \phi\}$ 
that will undergo adiabatic change. For future use we define 
\BEqA
&& A \equiv (1/2)(-B^2+6g^2)/\sqrt{B^4+4g^4}, \nonumber \\
&& A' \equiv (1/2)(B^2+6g^2)/\sqrt{B^4+4g^4}>0, \nonumber \\
&& K \equiv B^4-2g^2B^2+8g^4>0, \nonumber \\
&& L \equiv (-B^2+4g^2)\sqrt{B^4+4g^4}.
\EEqA

The corresponding Hamiltonian matrix
${\cal H}(B, \phi)$ has six different eigenvalues, two of which are doubly 
degenerate. 
 The non-degenerate ones are
$E_{5,6,7,8} = (1/2)[J\pm\sqrt{J^2 + 4(2g^2+B^2\pm 2Bg)}]$.
Their corresponding eigenkets are rather complicated, each leading to the same 
$U(1)$ Berry phase $e^{i\gamma_B} = e^{-3\pi i}$ on a single precession of the 
field, and will not be needed in what follows.
 The degenerate eigenvalues that {\it will} be needed are $ E_{3,4}=-J$, with
\BEq
 |\chi_{3}\ket = \frac{|001\ket-|010\ket}{\sqrt{2}}, |\chi_{4}\ket 
 = \frac{|110\ket-|101\ket}{\sqrt{2}},
\EEq
and $E_{1,2}=0$, with
\BEqA
\label{eq:STATES}
 |\chi_{1}(B, \phi)\ket &=& 
\{
(-B^2+4g^2)|000\ket
+B^2 e^{2i\phi}|011\ket \nonumber \\
&&-2gBe^{i\phi}|100\ket
\}/\sqrt{2K}, \nonumber \\
|\chi_{2}(B, \phi)\ket &=& 
\{
2gB^3e^{-3i\phi}|000\ket
-4 g^3B e^{-i\phi}|011\ket 
\nonumber \\
&& -B^2(B^2-2g^2)e^{-2i\phi}|100\ket
\nonumber \\
&&+K|111\ket
\}/\sqrt{2K(B^4+4g^4)}.
\EEqA
A straightforward calculation based on Eq. (\ref{eq:ADIABATICCONNECTION}) 
then gives the adiabatic connection
within the $|\chi_{1,2}\ket$ subspace,
\BEqA
d{\cal A}(B, \phi) &=&
i\{
[2B^2(B^2+2g^2)/K](\sigma^z/2) 
\nonumber \\
&& - (2gB^3A/K) \left(\sigma^x\cos 3\phi+\sigma^y \sin 3\phi\right) 
\}\; d\phi 
\nonumber \\
&& - i(2gB^2A'/K) ( \sigma^x \sin 3\phi - \sigma^y \cos 3\phi ) \; dB .
\nonumber \\
\EEqA
Here, the Pauli matrices operate within the $|\chi_{1,2}\ket$ ``geometric'' 
subspace and are different from the sigma-matrices originally used to 
describe the qubits in the circuit.

Notice that in the course of adiabatic evolution we are not allowed to take any 
of the limits $J\rightarrow 0$, $g\rightarrow 0$, or $B\rightarrow 2g$, since 
each  breaks the requirement of iso-degeneracy. 
Also notice that $J$ appears
nowhere in $|\chi_{1,2}\ket$, $E_{1,2}$, or $d{\cal A}$, so the geometric 
phase is not
sensitive to the precise value of $J$. Nevertheless, the presence of $J\neq 0$ is crucial here: it guarantees separation of 
the $|\chi_{1,2}\ket$ subspace from the rest of the Hilbert space.  Such (at least, partial) 
robustness against possible imperfections in the coupling is important since 
our ability to implement {\it multidimensional} geometric phases relies 
on the exact degeneracy of the underlying energy subspace. However, in 
order to have the exact degeneracy of $E_{1,2}$, we also require that the 
coupling $g$ be the
same for $1\leftrightarrow 3$ and $2\leftrightarrow 3$.  This is a very important 
requirement, not
easily  achievable experimentally (unless tunable coupling is available). 
Additionally, since $B$ {\it does} appear explicitly in Eq. (\ref{eq:STATES}), we
require good control over $B$, which is not a
problem experimentally.  In a typical experiment, even for long times, the 
magnetic field (the microwave drive) can easily be controlled \cite{NEELEY}.

We also point out that the sequence of operations generating non-abelian 
phases must be short compared to
the decoherence and dephasing times. On the other hand, system's evolution must remain 
adiabatic compared to the couplings $g$ and $J$ and the magnetic field $B$. 
In a typical experiment, $g$ is on the
order of 20 MHz, or $1/g = 50$ ns, so the entire operation would be several
hundred nanoseconds which is comparable to the presently achievable 
decoherence times. However, with stronger couplings up to $g = 100$ MHz and $1/g = 10$ ns
it is possible to perform an adiabatic operation in several tens of ns 
\cite{NEELEY}.

Let us now fix a basis $|\chi_{1,2}(B_0)\ket$ at $\phi = 0$ and initialize the 
system in some state 
$|\psi_0\ket = \a|\chi_{1}(B_0)\ket + \b |\chi_{2}(B_0)\ket$. We may then 
trace a closed loop ${\cal C}$ in the parameter space by performing the following 
sequence of operations:

\begin{figure*}
\includegraphics[angle=0,width=1.00\linewidth]{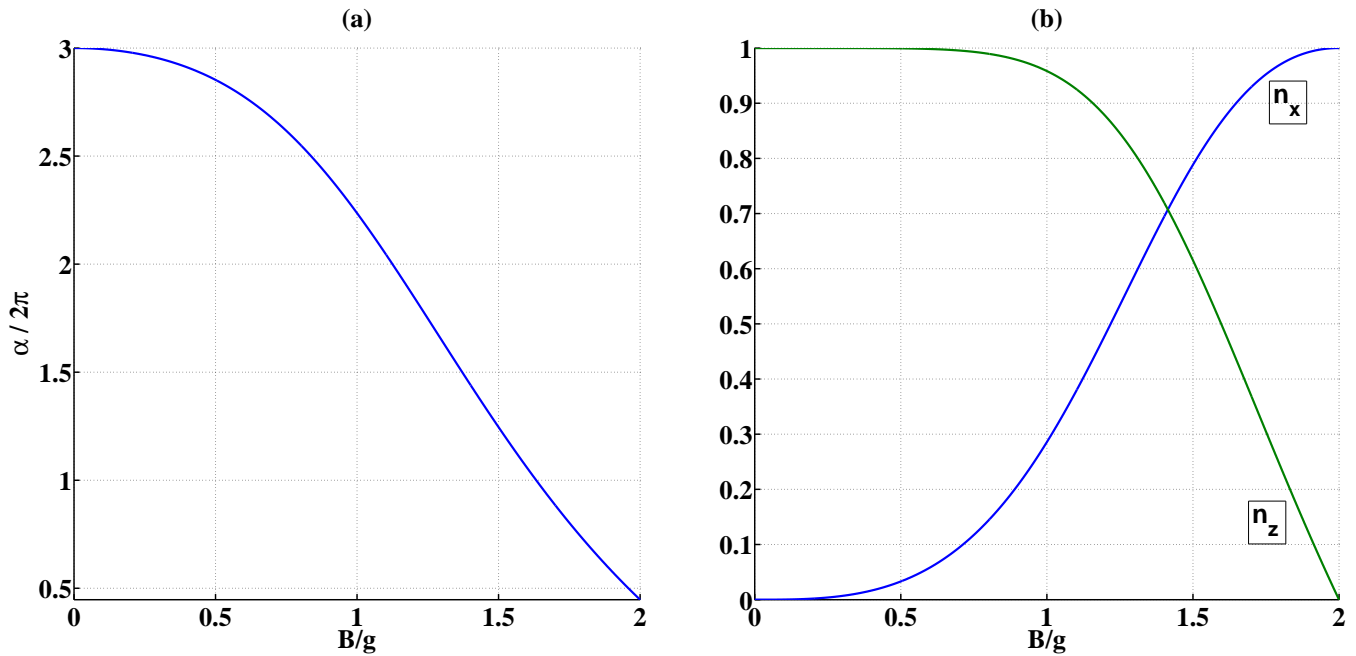}
\caption{
\label{fig:1} 
The one-parameter family of gates in the $|\chi_{1,2}(B)\ket$ 
basis that can be generated in a single precession using non-abelian holonomy 
$U_{XZ}(B)$ of Eq. (\ref{eq:U}) at $n=1$.
Panel (a): $B$-dependence of the rotation angle. Panel (b): $B$-dependence of 
the rotation axis. To implement a given gate (such as a Hadamard, NOT, PHASE, 
etc.), a specially 
designed computational basis, different from $|\chi_{1,2}(B)\ket$, must be chosen. 
See Sec. \ref{sec:COMPBASIS} for relevant examples.
}
\end{figure*}

(1) Begin, if needed, by adiabatically changing the field from $B_0$ to $B_1$ along 
the radial direction in the $(x,y)$-plane (no precession). The system will pick up a phase,
\BEqA
&& U_{Y} = e^{-i \theta (B_0,B_1) (\sigma^y/2)},
\nonumber \\
&&\theta(B_0,B_1) \equiv  2 \int_{B_0}^{B_1}  (2gB^2A'/K)\, dB .
\EEqA

(2) Perform adiabatic precession at $B_1$,
\BEq
B_{x}(t) = B_1 \cos \omega_{0}t, 
\quad 
B_{y}(t) = B_1 \sin \omega_{0}t .
\EEq 
If the field makes $n$ revolutions of total duration
$T_n = 2\pi n/\omega_{0}$, the accumulated phase will be 
\BEqA
\label{eq:U}
U_{XZ} &=& 
\underbrace{e^{-3i\pi n \sigma^z}}_{\mp 1}
\{ \cos(2A\pi n) 
\nonumber \\
&& + i \sin(2A\pi n) \left[(2gB^3/K)\sigma^x + (L/K)\sigma^z \right]\}{\arrowvert}_{B=B_1},
\nonumber \\
\EEqA
which represents a rotation by angle 
$\a_n=4A\pi n{\arrowvert}_{B=B_1}$ about the axis $\hat{n} 
= (-1)^{n-1}[2gB^3/K, 0, L/K]{\arrowvert}_{B=B_1}$ that lies in the $(x,z)$-plane 
of the Bloch sphere
(constructed with respect to $|\chi_{1,2}\ket$).
The set of gates that can be generated this way is shown in Fig. \ref{fig:1}.

(3) Bring the field back to its original value. This will result in additional contribution 
to the phase, $(U_{Y})^{-1}$. The full holonomy is then 
\BEq
\label{eq:generalHOLONOMY}
U_{{\cal C}} = (U_{Y})^{-1}U_{XZ}U_{Y} = U_{XZ} (U_{Y})^2.
\EEq

\section{Designing geometric gates}

\label{sec:COMPBASIS}

To demonstrate how this non-abelian scheme works in simple situations, let us 
find a basis in which a useful gate, such as a Hadamard, can be made in a single 
revolution of the effective magnetic field. We want $U_{Y}=1$, $n=1$, and 
$U_{{\cal C}} = U_{XZ} \equiv H$, possibly up to an insignificant U(1) factor. 
First, notice that the choice
$|\downarrow, \uparrow\ket = |\chi_{1,2}(B)\ket$ does not work as can be 
seen from Fig. \ref{fig:1}. We therefore introduce a new basis,
\BEqA
|\downarrow  \ket & :=& \cos \b |\chi_{1}(B)\ket   +  \sin \b |\chi_{2}(B)\ket , 
\nonumber \\
| \uparrow \ket &:=& - \sin \b|\chi_{1}(B)\ket  + \cos \b |\chi_{2}(B)\ket ,
\EEqA
with some unknown, $B$-dependent angle $\b$. Setting $n=1$ and changing 
to the new basis gives
\BEqA
\label{eq:U1Bbeta}
U_{XZ}(B,\beta) & = & -\cos(2A\pi) - i\sin(2A\pi) \times 
\nonumber \\
    &&   \{
    		 [ (L/K)\cos 2\b + (2gB^3/K)\sin 2\b ]\sigma^z \nonumber \\
	&&	 -[ (L/K)\sin 2\b - (2gB^3/K)\cos 2\b ]\sigma^x 
                            \}. \nonumber \\
\EEqA
It is then straightforward to check that by choosing
\BEqA
B&=& 
 g\sqrt{\frac{2\{-12+\sqrt{[39-4m_1(m_1+1)](1+2m_1)^2}\}}{4m_1(m_1+1)-3}},
  \nonumber \\
\beta_H &=& \arctan\left(\frac{1+(2gB^3/L)
+\sqrt{2\left[1+(2gB^3/L)^2\right]}}{1-(2gB^3/L)}\right)
\nonumber \\
&& -\frac{m_2\pi}{2},
\EEqA
with $m_1 \in  \{0,1,2\}$,  and thus $B/g \approx 1.9587$, $1.3716$, $0.8375$, 
and $m_2 \in \mathbb{Z}$, we get a Hadamard gate, $U_{XZ} = (\pm i) H$.
On the other hand, by choosing same $B$
and setting
$\b_{\rm NOT} =  \b_H - \pi/8$,
we get another basis, in which $U_{XZ} = (\pm i) {\rm NOT}$. 
Other important gates, such as $\sqrt{i\,{\rm NOT}}$ and various PHASE gates, 
can also be generated
in a similar manner.

We emphasize that the specially designed basis is only ``special'' in the sense
that it allows a given gate (say, a Hadamard) to be generated by following a 
particularly simple path in the parameter space --- here, by making a {\it single} 
turn of the adiabatically precessing in the $(x,y)$-plane field. 

\section{Conclusion}
\label{sec:CONCLUSION}

In summary, we have shown how capacitively coupled tripartite superconducting 
qubit system may support non-commuting SU(2) 
geometric holonomies. The non-abelian character of the holonomies is due to 
the inter-qubit coupling. If we treat the relevant degenerate subspace as a 
``geometric'' qubit, then, with a proper choice of the computational basis, any 
of the standard single-qubit gates used in quantum information processing can 
be generated by the adiabatic transport alone.

\begin{acknowledgments}

This work was supported by IARPA 
under grant W911NF-08-1-0336 and by the NSF under grant CMS- 
0404031. The author thanks Ken Brown, Michael Geller, and Matthew Neeley
for helpful discussions.

\end{acknowledgments}


\begin{thebibliography}{}

\bibitem{BERRY1984}
M. V. Berry, Proc. Roy. Soc. London A {\bf 392}, 45 (1984).
\bibitem{SIMON1983}
B. Simon, Phys. Rev. Lett. {\bf 51}, 2167 (1983).
\bibitem{WILCZEK&ZEE1984}
F. Wilczek and A. Zee, Phys. Rev. Lett. {\bf 52}, 2111 (1984).


\bibitem{geomPhaseLiterature}
See, e.g., A. Tomita, R. Y. Chiao, Phys. Rev. Lett. {\bf 57}, 937 (1986);
C. A. Mead, Rev. Mod. Phys. {\bf 64}, 51 (1992);
D. Arovas, J. R. Schrieffer, and F. Wilczek, Phys. Rev. Lett. 53, 722 (1984);
H. Koizumi, T. Hotta, Y. Takada, Phys. Rev. Lett. {\bf 80}, 4518 (1998).

\bibitem{geomPhaseBooks}
A. Shapere, F. Wilczek, {\it Geometric Phases in Physics}, World Scientific (1989);
A. Bohm, A. Mostafazadeh, H. Koizumi, Q. Niu, J. Zwanziger, {\it The Geometric 
Phase in Quantum Systems},
Springer (2003).

\bibitem{HQC}
P. Zanardi, M. Rasetti, Phys. Lett. A {\bf 264}, 94 (1999); 
J. Pachos, P. Zanardi, Int. J. Mod. Phys. B {\bf 15}, 1257 (2001);
A. Eckert, et. al., J. Mod. Opt. {\bf 47}, 2501 (2000).

\bibitem{nonAbelianScenarios}
L. Faoro, J. Siewert, and R. Fazio, Phys. Rev. Lett. {\bf 90}, 028301 (2003);
M. Cholascinski, Phys. Rev. Lett. {\bf 94}, 067004 (2005);   
V. Brosco, R. Fazio, F. W. J. Hekking, and A. Joye,
Phys. Rev. Lett. {\bf 100}, 027002 (2008).

\bibitem{YOU_review05}
See, e.g., 
J. Q. You, F. Nori, Phys. Today {\bf 58} (11), 42 (2005);
G. Wendin and V. S. Shumeiko, Low Temp. Phys. {\bf 33}, 724 (2007).

\bibitem{MARTINIS2007}
M. Steffen, M. Ansmann, R. C. Bialczak, N. Katz, E. Lucero, R. McDermott, 
M. Neeley, E. M. Weig, A. N. Cleland, J. M. Martinis, Science {\bf 313}, 1423 (2006).

\bibitem{maxEnt2008}
A. Galiautdinov, J. M. Martinis, Phys. Rev. A {\bf 78}, 010305(R) (2008).

\bibitem{NEELEY}
M. Neeley (private communication).

\end{thebibliography}
\end{document}